\documentclass[twocolumn,prb,aps]{revtex4-1}
\usepackage{graphicx}% Include figure files
\usepackage{color}
\usepackage{dcolumn}% Align table columns on decimal point
\usepackage{bm}% bold math
\usepackage{float}
\usepackage{hyperref}%
\usepackage{float}

\begin{document}
\title{Defect-induced Fermi level pinning and suppression of ambipolar behaviour in graphene}
\author{Zakaria Moktadir}
\affiliation{Electronics and Computer Science, Faculty of Applied Physical
Sciences, Southampton University, United Kingdom}
\email{zm@ecs.soton.ac.uk}
\author{Shuojin Hang}
\affiliation{Electronics and Computer Science, Faculty of Applied Physical
Sciences, Southampton University, United Kingdom}
\author{Hiroshi Mizuta}
\affiliation{Electronics and Computer Science, Faculty of Applied Physical
Sciences, Southampton University, United Kingdom}
\affiliation{School of Materials Science, JAIST, Nomi, Ishikawa 923-1292, Japan}

%\pacs{72.10.Fk, 73.63.-b, 81.05.Uw }

%%%%%%%%%%%%%%%%%%%%%%%%%%%%%%%%%%%%%%%%%%%%%%%%%%%%%%%%%%%%%%%%%%%%%
%% The document title should be given as usual
%% A short title can be given as a *suggestion* for running headers.
%%%%%%%%%%%%%%%%%%%%%%%%%%%%%%%%%%%%%%%%%%%%%%%%%%%%%%%%%%%%%%%%%%%%%

%\keywords{Graphene, impurity, scattering, defects, edge scattering}

%%%%%%%%%%%%%%%%%%%%%%%%%%%%%%%%%%%%%%%%%%%%%%%%%%%%%%%%%%%%%%%%%%%%%
%% The manuscript does not need to include \maketitle, which is
%% executed automatically.  The document should begin with an
%% abstract, if appropriate.  If one is given and should not be, the
%% contents will be gobbled.
%%%%%%%%%%%%%%%%%%%%%%%%%%%%%%%%%%%%%%%%%%%%%%%%%%%%%%%%%%%%%%%%%%%%%
\begin{abstract}
We report on systematic study of electronic transport in low-biased, disordered graphene nanowires. We reveal the emergence of unipolar transport as the defect concentration increases beyond 0.3\% where an almost insulating behaviour is observed on n-type channels whilst a metallic behaviour is observed in p-type channels. The conductance shows a plateau that extends through the entire side above the Dirac point (n-type) and the conductivity coincides with the minimum conductivity at the Dirac point.  The minimum conductivity decreases with increasing defect concentration pointing out towards the absence of zero energy modes in the disordered samples. Raman spectroscopy and X-ray photoemission spectroscopy were used to probe the nature of the defects created by helium ion irradiation and revealed the presence of oxygen-carbon bonds as well as the presence of $sp^3$ configuration uncovered from the C KLL Auger spectrum. The observed behaviour is attributed to the dangling bonds created by sputtering of carbon atoms in graphene lattice by bombarding helium ions. The dangling bonds act as charge traps, pinning the Fermi level to the Dirac point.
\end{abstract}

\maketitle
\section{Introduction}
Graphene, a truly 2-dimentional material\cite{Novoselov04}, has attracted enormous interest in the last decade and is listed as one of the major materials for post-CMOS era, according to the International Technology Roadmap for Semiconductors  \cite{ITRS}.
However, the absence of a band gap in graphene is a major hurdle towards its utilization in logic circuits for applications in electronics.  Although the conductivity of graphene is tunable using electrostatic gating, the existence of a minimum conductivity \cite{Ten07,Adam07,Sarma2011} results in a finite "off" current and  consequently a small on-off ratio is observed. Researchers have developed a wide variety of functionalization methods to open a band gap in graphene. These modification techniques comprise fluorization\cite{Withers11,Hong11,Leenaerts10,Nair10,Cheng10,Withers10,Lee12}, hydrogenation \cite{Elias09,Balog10,Lu09,Choe10,Chuang12,Sofo07,Ryu08,Choe12,Matis12,Havu11,
Bostwick09,Haberer2010,Bang10}, oxidation\cite{Masubuchi11,Seo13,Wu08,Guo2012,Navaro07,Fujii10,Yan2009}, chlorination\cite{Sahin12,Li2011,Xu2013,Vinogradov12,Yang2012,Ijas2012,Wu2011} or functionalization with other chemical species\cite{Englert2011,Usachov2011,Liu2012,Hirsch2012,Sun2010,Georgakilas2012,Niyogi2010,
Joucken2012,Zhang2011}. Such functionalization has the effect of modifying the local electronic structure of   carbon atoms in the hexagonal lattice, and functionalized sites can act as short range scatterers by modulating the local potential. The resulting random potential distribution in graphene sheet may induce localization phenomena\cite{Suzuura2003,McCann2006,Louis2007,Lherbier2008_1,Aleiner2006,Gunlycke2007,Ponomarenko2001,
Evaldsson2008,Nakaharai2013,Kotakoski2011,Tuan2012,Zhou2010,Guangyu_Xu2011} or it can result in the emergence of a band gap \cite{Balog10,Haberer2010,Zhang2011}.   Quasi-1D systems such as graphene nanoribbons or nanowires are of particular interest where edges strongly affect electronic states as they play a major role in the localization of carriers \cite{Guangyu_Xu2011}. Beside the role played by edges, the localization of 2D electronic states can also occur by the deliberate introduction of structural defects in bulk graphene. A large amount of defects can be introduced in graphene by electron or ion bombardment for instance. A high level of disorder induced by electron irradiation was shown to possibly result in the formation of a two-dimensional amorphous carbon lattice\cite{Kotakoski2011}. Gallium ion irradiation on graphene flakes showed a structural transition from nano-crystalline graphene to amorphous carbon where the charge transport is dominated by variable range hoping process and where very short mean free path and localization lengths were observed \cite{Zhou2010}.
In a recent report\cite{Nakaharai2013}, it was shown  that the introduction of defects in graphene narrow channels using lighter ions such as helium ions can be used to significantly alter graphene electronic properties. An  on-off current ratio of 100 was found at room temperature.  Scaling of the conductance was also observed showing an exponential decay as a function of the irradiated channel length implying that the transport is dominated by strong localization even at room temperature. This contrasts with the finding of reference \citenum{Guangyu_Xu2011} where edge disorder in narrow channels is responsible for strong localization.\\ 
In this paper we investigate transport properties of disordered graphene field effect nanowire devices.  We found that as the disorder is increased, the Fermi level is pinned to the Dirac point and the ambipolar characteristic of graphene is suppressed whilst the conductance shows a plateau that extends through the entire side above the Dirac point (n-type) and the conductivity coincides with the minimum conductivity at the Dirac point.  The minimum conductivity decreases with defect concentration pointing out towards the absence of zero energy modes in the defective samples.  Despite the small on-off current ratio, the suppression of the ambipolar behaviour using helium ion irradiation is an essential requirement towards using irradiated graphene FETs in electronics.

\section{Results and discussion}
\begin{figure*}[ht]
\includegraphics{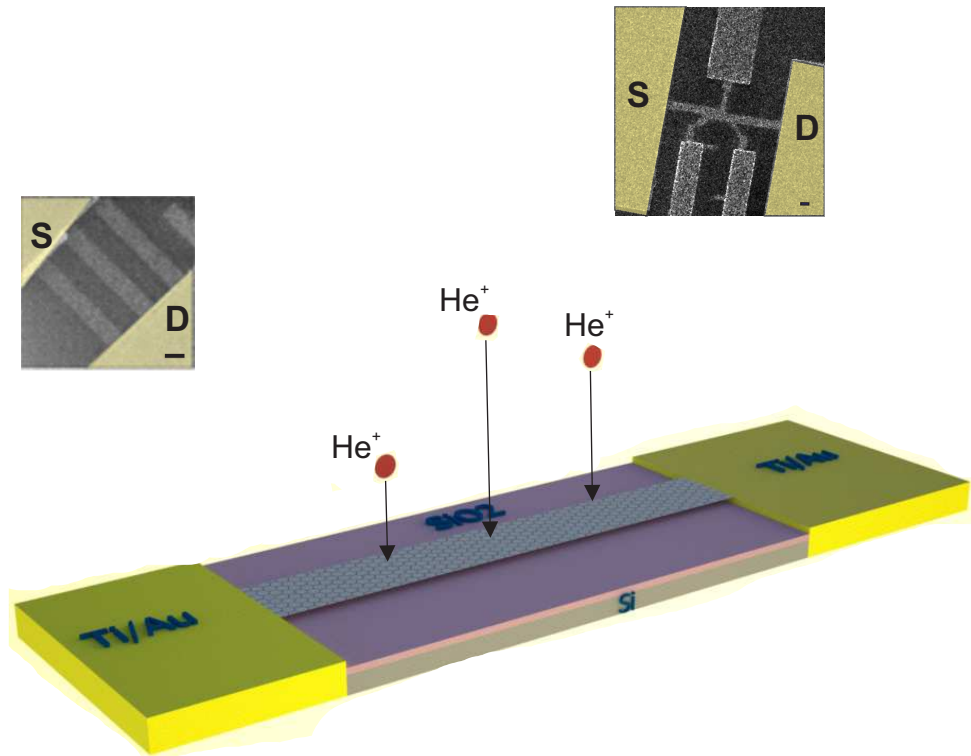}
\caption{(a) Main panel: sketch of the fabricated device showing the channel, source and drain contacts. The channel is precisely irradiated with helium ions. The silicon substrate is used as a gate electrode whilst the silicon oxide layer ($\sim 300 nm$) is used as a gate dielectric. Inset: the helium ion microscope micrographs of the fabricated two and multi-terminal devices used in this work, showing source, drain contacts and voltage probes. The scale bar is 200 nm for both insets.}
\label{fig1}
\end{figure*}
\begin{figure}
  \includegraphics[scale=.45]{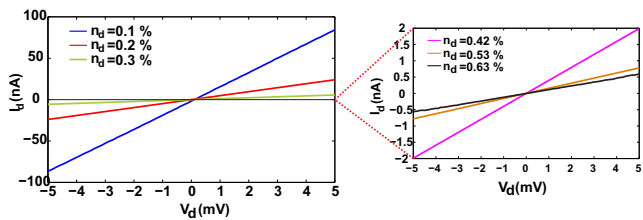}
  \caption{(a) Drain current-voltage characteristics of an irradiated channels which were 200 nm wide and 1 micron long. These characteristics were measured at different values of defect concentration. One immediately notices the absence of a transport gap around zero-bias.}
\label{fig2}
\end{figure}

In figure \ref{fig1} a sketch of the device is shown alongside a helium ion microscope (HIM) micrographs of typical two-terminal and multi-terminal devices considered here. Only the channel areas was subjected to irradiation with an accuracy less than 2 nm. The source and the drain contacts were not irradiated to avoid any variations in the contact resistance. In figure \ref{fig2}-a the measured I$_d$-V$_d$ characteristics for six values of defect concentration $n_d$ ($n_d$ = 0.1 \%-0.63 \%) in  a two-terminal device are shown.  The I$_d$-V$_d$ characteristics are linear for both, pristine and irradiated devices and show a clear absence of a transport gap in contrast with the findings in reference \citenum{Nakaharai2013} where a transport gap was clearly observed in channels with similar width (200 nm) but shorter two terminal devices, at room temperature.   The resistivity of the device increases with the defect concentration indicating an enhancement of carrier scattering by the created point defects.  The defects were revealed by Raman spectroscopy which shows an increase in the D-peak as the irradiation dose is increased (see supporting information). 

\begin{figure}
\includegraphics[scale=0.5]{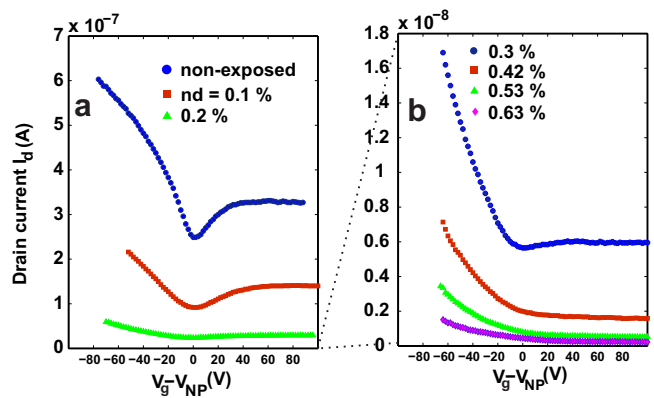}
\caption{a) I$_d$-V$_g$ curves for non-irradiated and moderately irradiated graphene nanoribbons. b) I$_d$-V$_g$ curves for heavily disordered graphene ($n_d \geq 0.3\%$. Here the drain voltage is V$_d$= 5 mV. }
\label{I_Vg}
\end{figure}

\begin{figure}
\includegraphics[scale=0.5]{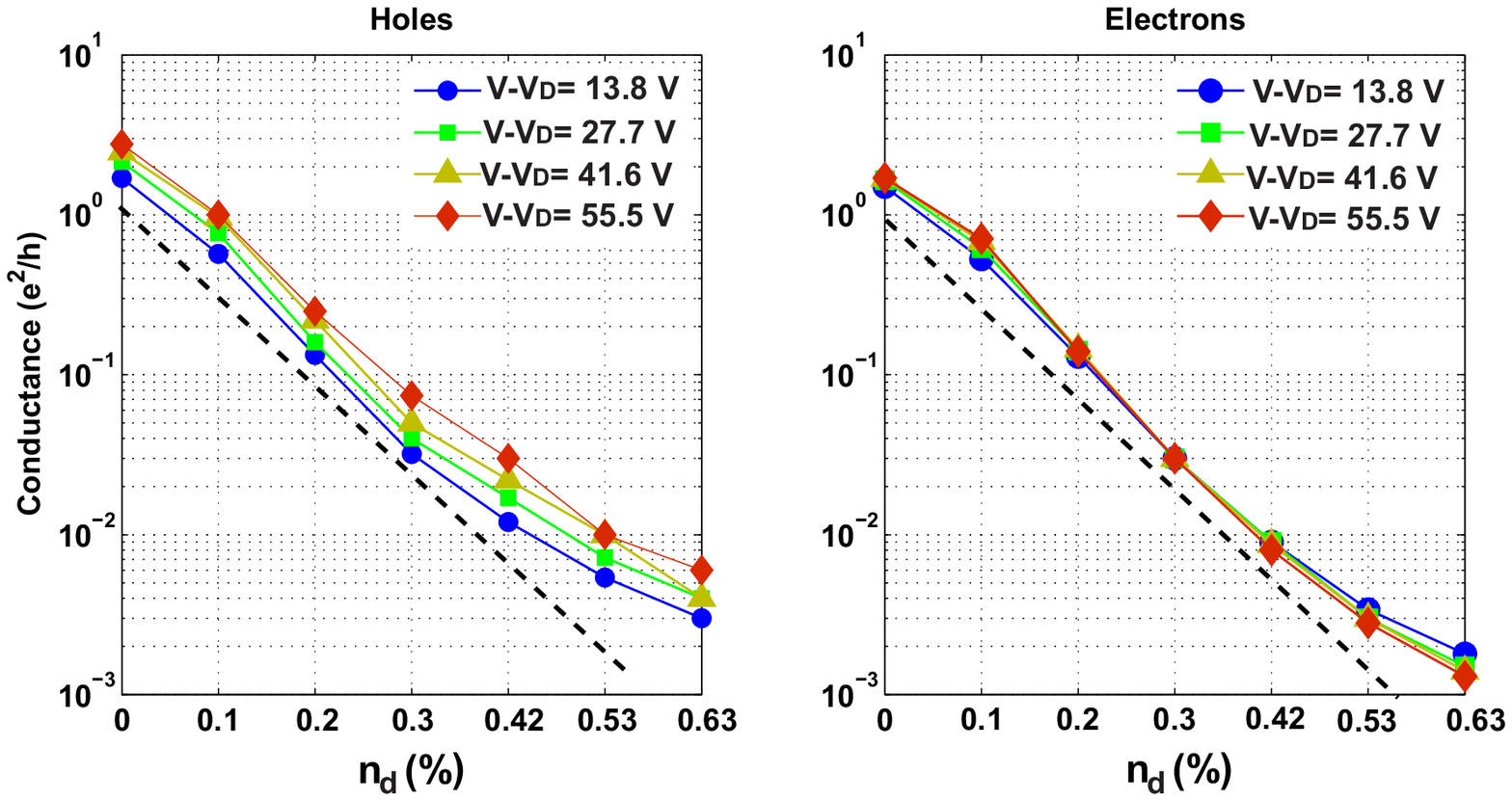}
\caption{The exponential decrease of the conductance of a two-terminal device as a function of the defect density for holes and electrons.}
\label{G_vs_dose}
\end{figure}

\begin{figure*}
\centering
\includegraphics[scale=0.5]{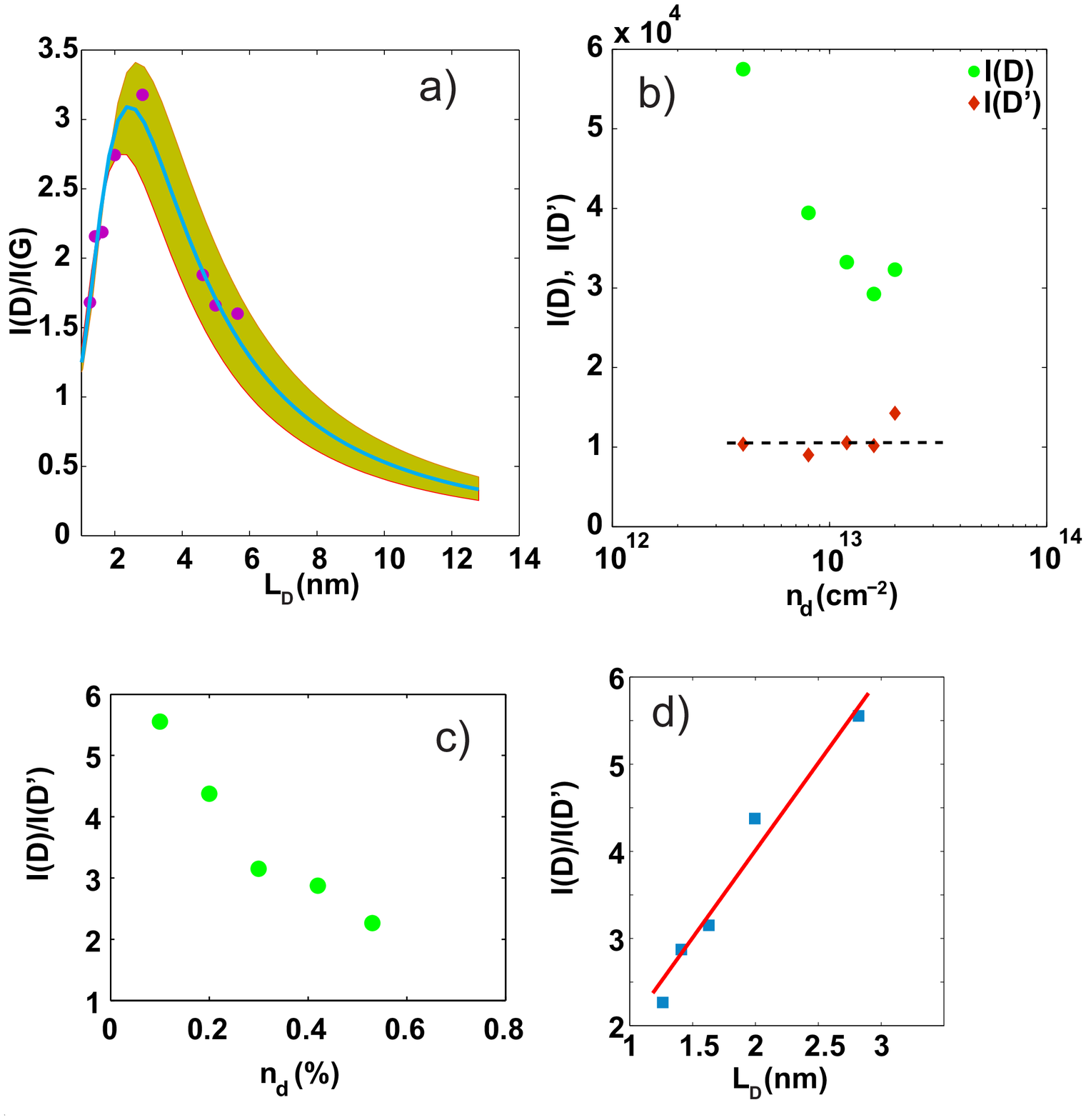}
\caption{a) Plot of the ratio I(D)/(G) as a function of the average distance between defects $L_D$. The circles correspond to experimental data whilst the continuous bleu line corresponds to the best fit of the data to equation \ref{activation}. The shaded area shows the uncertainty range corresponding to the parameters $C_A$, $r_A$, $r_S$ and $C_s$ determined form the fitting procedure. In b) the plot of Raman intensities I(D) and I(D'), and in c) the ratio I(D)/I(D') are shown as a function of the defects concentration. In d) I(D)/I(D') is plotted versus the average distance between defects.}
\label{activation2}
\end{figure*}

In figure  \ref{I_Vg}-a we show the plot of the channel current versus the gate voltage (I$_d$-V$_g$ curves) for non-irradiated and moderately irradiated graphene channels. In these plots the usual graphene characteristic is preserved showing an ambipolar behaviour and a neutrality point (NP). The conductivity varies linearly with the carrier concentration further away from the NP and shows a sub-linear behaviour at high values of $V_g$ due to short-range scattering \cite{Sarma2011}. As expected, a decrease in the drain current is observed as the defect concentration is increased above 0.3\%.  Above this value the characteristics show a striking behaviour (figure  \ref{I_Vg}-b): the devices no longer display the usual graphene I$_d$-V$_g$ curves but rather a flat characteristic above the NP (n-type transport) and a superlinear behaviour below the NP (p-type transport).
In figure \ref{G_vs_dose} the conductance $I_d/V_d$(in units of $e^2/h$) as a function of the defect concentration and the gate voltage is shown. From the semi-log plot we notice that the conductance decreases exponentially with the defect concentration.
We performed the same measurements on multi-terminal devices (see supporting information) and similar observations apply, that is the suppression of the ambipolar characteristic in irradiated channels. These multi-terminal devices were chosen to be wider (i.e. $w=1 \mu m$), to avoid the influence of the edges and thus assert that the observed behaviour is purely due to bulk defects created by He$^+$ bombardment.\\
\indent To gain more understanding of the observed transport behavior and its correlation to the microscopic structure of disordered graphene, we performed Raman spectroscopy on irradiated graphene samples for similar defect concentrations.  Before irradiation the D-peak which is indicative of disorder, is absent (see supporting information). As the defect concentration is increasing the prominence of D-peak is enhanced (at $\sim$ 1350 cm$^{-1}$), as expected. The ratio of intensities for the D-band and the G-band decreases with the defect concentration.  In terms of the average distance between defects $L_D=\sqrt{{10^{14}/\pi n_d}}$, this ratio is increasing for $L_D \leq 3$ nm, indicating a stage 2 disorder as shown in figure \ref{activation2}-a.  The fit to the activation model\cite{Lucchese2011}  is also shown . The activation model predicts the following expression for the ration I(D)/I(G):
\begin{equation}
\frac{I(D)}{I(G)}=C_A\frac{(r_A^2-r_S^2)}{(r_A^2-2r_S^2)}\left(e^{{\frac{-\pi r_S^2 }{L_D^2}}}-e^{\frac{-\pi (r_A^2-r_S^2)}{L_D^2}}\right) \\
+C_s\left(1-e^{{\frac{-\pi r_S^2 }{L_D^2}}}\right)
\label{activation}
\end{equation}

\begin{figure}
\includegraphics[scale=0.45]{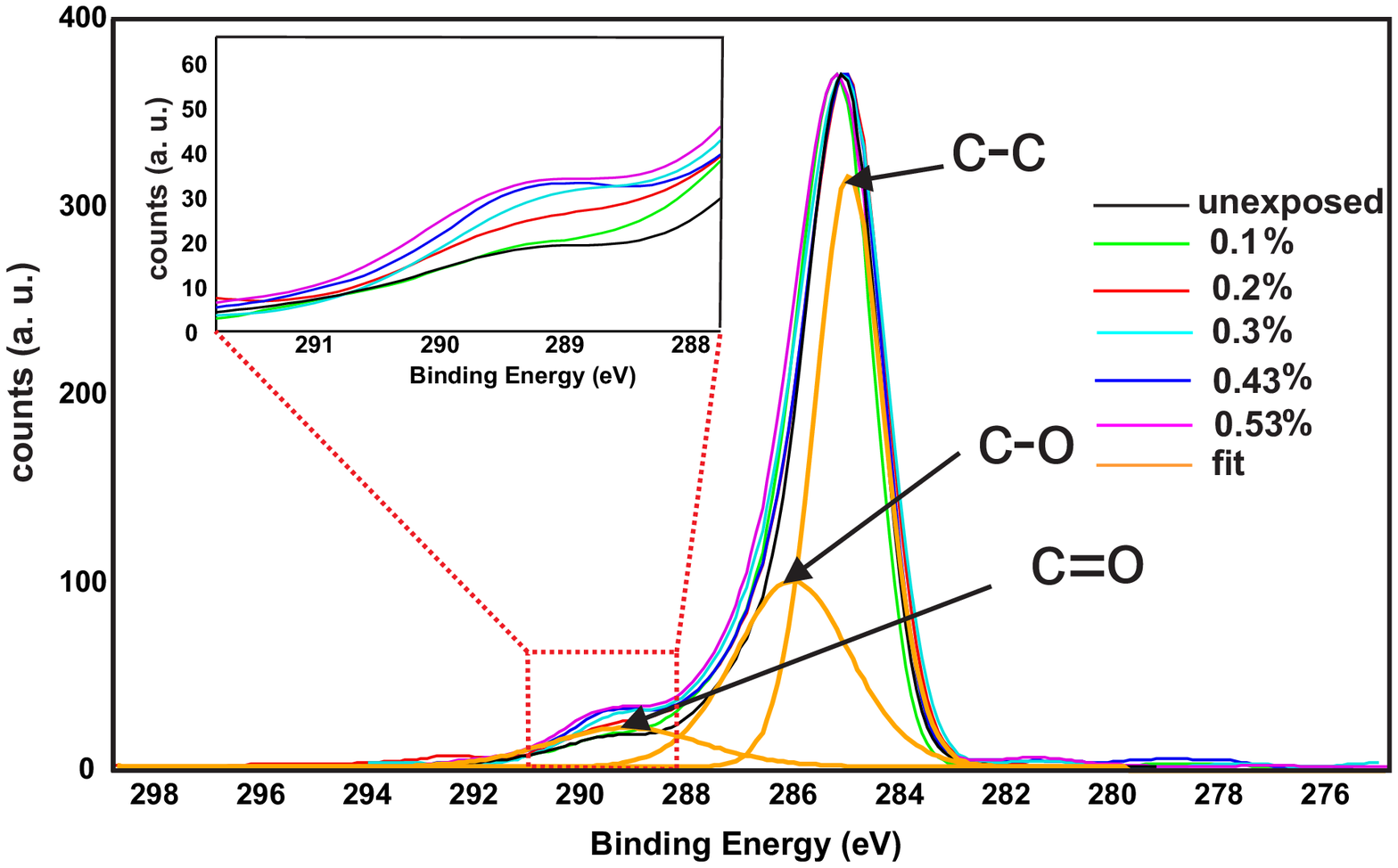}
\caption{High resolution XPS spectrum of C1s core level for different values of defect concentration as indicated. The fit to one of the spectra (orange curve) shows the presence of three peaks at 284.4 eV, 286.5 eV and 288 eV. The inset shows the zoom-in of the area of the spectrum situated around the third component.}
\label{XPS}
\end{figure}

To increase the accuracy of the fitting procedure we added data for doses below 1 $\times$ 10$^{15}$ cm$^{-2}$ or for $L_D$ $\geq$ 3 nm.  The parameter $r_s$ corresponds to the radius of the structurally disordered area surrounding the impact point of an incident helium ion, whilst $r_A$ is the radius of the area where the D-band transition occurs.  The fit to equation \ref{activation} gives the following values for the parameters $C_A=4.22 \pm 0.6$, $r_S= 0.96 \pm 0.13$ nm, $r_A=2.24 \pm 0.16$ nm and $C_S=0.97 \pm 0.05$.The Raman relaxation length for the resonant Raman scattering $l=r_A-r_S=1.26 \pm 0.05 nm$, which is consistent with the finding in reference \citenum{Lucchese2011} for graphene irradiated with Ar$^+$. The parameter $C_A$ corresponds to the maximum value of the ratio $I_D/I_G$ and its value derived here is  smaller than the value found for graphene irradiated with Ar$^+$ for the same excitation wavelength\cite{Cancado2011}.\\ 
\indent An equally important quantity which is useful to probe the nature of defects is the ratio I(D)/I(D') \cite{Eckmann2012}. In figure \ref{activation2}-b Raman intensities I(D) and I(D') (the D' peak is at 1600 cm$^{-1}$) are plotted as a function defect concentration $n_d$. The D-band intensity decreases with $n_d$ whilst the D'-band is roughly constant for the range of $n_d$ present in our samples. As a consequence the ratio I(D)/I(D') decreases with defect concentration as shown in figure \ref{activation2}-c. In stage 1, this ratio should not depend on the defect concentration but will only depend on the nature of defects\cite{Eckmann2013}. The data shown in figure \ref{activation2} is a clear indication that our irradiated samples are in stage 2 disorder.
The decrease in the D-band intensity is consistent with theoretical predictions \cite{Venezuela2011} where the maximum of I(D) is reached for a concentration of 6 $\times$ 10$^{12}$ cm$^{-2}$ before it starts decreasing for higher defects concentrations. The decrease in I(D) happens when the average length an electron/hole travels in between two scattering events with a defect becomes smaller than the average length an electron/hole couple travels before scattering with an optical phonon \cite{Venezuela2011}. In this case, sp$^2$ groups become reduced and the rings become distorted. The strength of the D-peak is found to vary linearly with the average distance between defects according to the relation $I(D) \approx \alpha L_D \propto 1/\sqrt{n_d}$ (figure \ref{activation2}-c,d). Consequently, I(D)/I(D') $\propto$ $1/\sqrt{n_d}$ as I(D') is roughly constant over the range of defect concentrations present in our samples.\\    
Contrary to stage 1 disorder where a correlation between the nature of defects and the ratio I(D)/I(D') can be drawn; in stage 2 however,  it is still not clear how to relate the defects nature to the intensity ratio I(D)/I(D'). More systematic study is needed to clarify this correlation.\\
\indent Another precise analytical method used to quantify disorder in graphene is the X-ray photoemission spectroscopy (XPS). In this work we performed XPS on irradiated graphene samples (for details see supporting information).  First, a survey in the entire range of binding energy $[$0, 1200 $]$ eV was performed (see supporting information) to identify the different atomic species present in our samples besides carbon. No other elements were found apart from oxygen and silicon originating from the underlying SiO$_2$, although some oxygen may be present in the graphene sheet forming bonds with carbon. Figure \ref{XPS} shows the C1s core level spectrum for different values of defect concentration.
  
\begin{figure}
\includegraphics[scale=0.45]{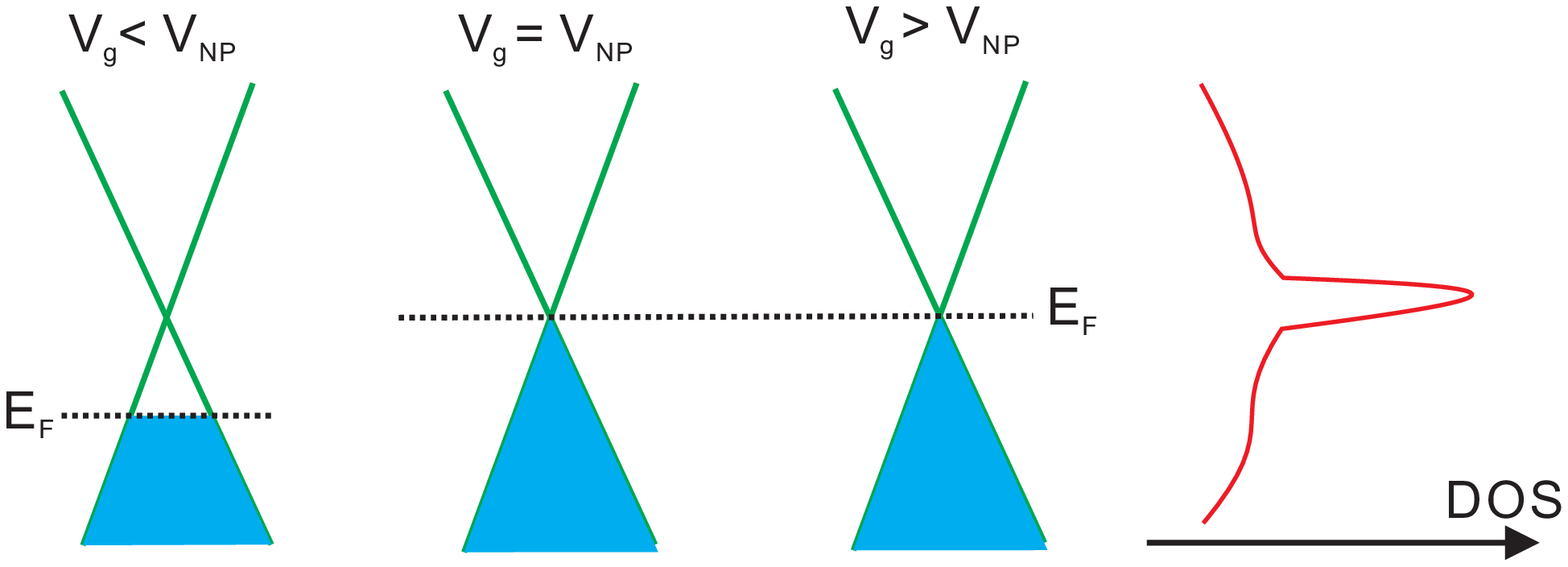}
\caption{Sketch of the Fermi level pinning at the Dirac point. When the gate voltage is higher than the Dirac voltage, the conductivity is aligned with the minimum conductivity which determined by the defect-induced peak of the density of states (also shown) at the Fermi level. When the gate voltage is below the Dirac voltage, the Fermi level is unpinned and can freely move inside the valence band giving rise to hole conduction which changes with the Fermi level.}
\label{fermi_level}
\end{figure}

All the spectra show an asymmetric broad line and can be split into three components. The spectra were fitted to a convolution of Gaussian and Lorentzian functions. The major spectral component shows a slight shift in binding energy compared to the sp$^2$ line in graphite which is 284.4 eV  \cite{Speranza07,Emstev2008}. This shift might be attributed to charge transfer between the graphene sheet and the underlying oxide substrate. It was shown from previous works that the measured C1s core level binding energy in graphene depends on the underlying substrate and the difference can be few hundreds of meV \cite{Haberer, Zhou03, Lizzit}. Therefore we use the value for graphite i.e. 284.4 eV as a reference in all spectra measured for different defect concentration. The FWHM of the peak at 284.4 eV increases with defect concentration indicating broadening. The sp$^2$  component persists even for the highest defect concentration $n_d=0.63 \%$, meaning that the graphene atomic structure is still preserved to a  significant degree. For different defects concentrations, the other two components are roughly centered at 286 eV and 288.5 eV. These two peaks can be attributed to C-O and C=O  groups                                                                 respectively\cite{Dreyer,Hong2012,Islam2011}. Similar observation were found in few layer graphene irradiated with electron \cite{Xu2011} and irradiated graphite \cite{Speranza07}. The sp$^3$ to sp$^2$ contents can be determined by the so-called the D-parameter which measures the distance between the maximum and the minimum in the carbon (C KLL) Auger electron first derivative\cite{Jackson95, Mezi2010} spectrum. For pure graphite (pure sp$^2$) this parameter is 21 eV whilst it has a value of 13 eV diamond (pure sp$^3$)\cite{Mezi2010}. There is a linear relationship between the sp$^2$ to sp$^3$ ratio and the value of D\cite{Mezi2010}. We found a value of 18 eV for our He$^+$ irradiated graphene samples which did not vary significantly in the range of the defect concentrations used (0.1\% - 0.63\%). This indicated a ratio  of sp$^2$ to sp$^3$ contents of 0.6 (see supporting information).\\  Helium ion sputtering of carbon atoms induces the presence of vacancies of different nature, although our XPS measurements couldn't resolve the location of peaks corresponding to vacancies due to the limited resolution of our XPS instrument. It was shown using density functional theory that the energy shift between the C1s peak location in graphene and the vacancies peaks  is -0.64 eV, -0.49 eV and -0.34 eV for single vacancy, di-vacancy and Stone-Wales vacancy respectively\cite{Susi2014}.These shifts were not resolved in our XPS spectrum. On the other hand, Oxygen groups and vacancies act as short ranges scattereres and their presence induce a peak in the density of states at the Fermi level\cite{F_Gui08,Weh09,Weh08}.  This particular type of defects is also called resonant scatterers as they strongly couple to graphene electronic states and thus scatter carriers very efficiently \cite{Weh09_2}. Covalently bonded adatoms and vacancies can also give rise to zero energy modes\cite{Pereira08} (ZEMs) which produce a supermetallic regime by enhancing the Dirac-point conductivity above its minimum
ballistic value $\sigma_{min}=4e^2/\pi \hbar$\cite{Cresti2013}. The deceasing minimum conductivity in our samples (the conductivity is measured in multiprobe samples, see supporting information) with increasing defects concentration points out towards the absence of ZEMs\cite{Cresti2013}. For one of the samples the conductivity
reaches the semiclassical limit of minimum conductivity\cite{Lherbier12} at $n_d=0.3\%$. The present result is consistent with semi-classical Boltzmann transport theory and suggests the absence of quantum interference and strong localization.\\   The observed unipolar character of the conductance and the electron-hole asymmetry is very reminiscent to that predicted theoretically for H$^+$ adsorbates, whilst for OH$^-$ absorbate the conductance shows an n-type behaviour\cite{Robinson09}. However it is not clear if water molecules from the ambient provide the proton and the hydroxide group to an unsaturated dangling bond. This can be potentially probed by in-vacuum hydrogenation of irradiated graphene which has a passivation effect. More research is needed to clarify this.  The presence of $sp^3$ structures revealed from the D-parameter can be attributed to out-of-plane deformation induced by defects as predicted theoretically for Stone-Wales defects\cite{Ma09} or by adsorbates that re-hybridize the sp$^2$ orbitals. In the present case, the main adsobates revealed by XPS are oxygen which form single and double bonds with carbon atoms.  
When an incident He$^+$ ion dislodges a carbon atom from the honeycomb lattice, it leaves three dangling bonds; two of them can recombine to form a double bond leaving one unsaturated dangling bond. The flatness of the conductance on our n-type samples can be attributed to the existence of unsaturated dangling bonds which act as charge traps. These traps are associated with a large density of states at the Dirac point which pins the Fermi level. Below the Dirac point the Fermi level is free to move within the valence band triggering hole current as sketched in figure \ref{fermi_level}. For large defect concentration, the conductivity in the n-type channel coincides with the minimum conductivity and the minimum conductivity plateau usually seen in graphene extends to the whole n-side.

\section{SUMMARY AND CONCLUSIONS}
In conclusion we have systematically investigated the transport properties of field effect graphene devices irradiated with accelerated helium ions. The defect concentration was controlled by increasing the dose of the irradiated ions, and was in the range 0.2\%-0.63\%. As the defect concentration increases, an almost insulating behaviour is observed in n-type graphene whilst p-type graphene shows a metallic behaviour. The conductance shows a plateau that extends through the whole side above the Dirac point (n-type) . This was a result of Fermi level pinning at the Dirac point which was induced by unsaturated dangling bonds acting as charge traps. Raman spectroscopy and X-ray photoemission spectroscopy were used to probe the nature of the defects created by helium ion irradiation and revealed the presence of oxygen-carbon bonds as well as $sp^3$ configuration on the irradiated samples. The observed unipolar behaviour is an important step towards using graphene in logic circuits. However the question remains how to increase the on-current on the p-side and equally how to create an insulating behaviour on p-type side which will allow complimentary logic operations. One solution is to decrease the channel length; this is an ongoing investigation.

\section*{METHODS}
The graphene used in this work is obtained from mechanical exfoliation of highly oriented pyrolytic graphite (HOPG) on a 300 nm SiO2 layer sitting on a highly doped silicon substrate  which is used as a back gate terminal. Single layer flakes were identified using both an optical contrast method and Raman spectroscopy \cite{Hiura2010,Ferrari2006}. The Graphene nanowires were defined using ebeam lithography and oxygen plasma etching.  This is followed by a second lithographic layer to deposit thermally evaporated  Ti/Au (10 nm/80 nm) contacts using ebeam lithography and a lift-off process using a double layer of  methylmethacrylate (MMA) and poly(methylmethacrylate) (PMMA) resists. The fabricated devices were baked overnight at 250 C to free the channels from contamination caused by the resist residue. The irradiation with helium ions He$^+$ was performed in high vacuum inside a helium ion microscope (Zeiss Orion) \cite{Moktadir2011,Lemme2009,Bell2009,Fox2013,Winston2009}. The acceleration voltage used was 30kV whilst the current was kept at 1 pA. To avoid variations of contact resistance, contacts were not exposed to the He$+$ beam and only graphene channels were irradiated. This was possible by the high precision patterning offered by the helium ion microscope, with a He$^+$ beam spot size of 0.7 nm.  The channels were perfectly mapped using an embedded pattern generator. Experiments were performed on both two-terminal and multi-terminal devices. The measurement were performed immediately after each irradiation run, using an Agilent B1500 semiconductor analyser.  The channels were exposed to doses of $n \times 10^{15} cm^{-2}$ with n =1,2,3,4,5 and 6. The estimated defect concentration corresponding to these doses is $n_d$=0.1 \%, 0.2\%, 0.3\%, 0.42\%, 0.53\% and 0.63\%\cite{Nakaharai2013,Bell2009} respectively. \\
Raman spectroscopy (Renishaw inVia ) was performed on samples using freshly exfoliated graphene flakes which were subjected to the same doses as above. The laser excitation wavelength used was 532 nm (i.e. 2.3 eV).
The XPS was performed on a freshly transferred CVD graphene onto Si/SiO$_2$ substrate (see supporting information), using a Theta Probe Angle-Resolved X-ray Photoelectron Spectrometer, from Thermo Scientific$^{TM}$. The system uses a Monochromated, Micro-focused Al K-Alpha source delivering a spot size between 15 $\mu m$ and 400 $\mu m$. For our samples a spot size of 100 $\mu m$ was chosen for optimum measurements. \\
%\indent {\it Conflict of Interest}: The authors declare no competing
%financial interest.

%\section*{ REFERENCES AND NOTES}

\end{document}